\pdfoutput=1

\documentclass[11pt,twoside,a4paper,cmspaper,final,collab]{cms-tdr}
\def\svnVersion{446985P}\def\cmsCernNoTag{CERN-EP-2017-262}\def\cmsCernDate{\today}\def\cmsMessage{Published in the European Physical Journal C as \href{http://dx.doi.org/10.1140/epjc/s10052-018-5607-5}{\doi{10.1140/epjc/s10052-018-5607-5.}}}

\begin{document}\cmsNoteHeader{TOP-17-009}

\hyphenation{had-ron-i-za-tion}
\hyphenation{cal-or-i-me-ter}
\hyphenation{de-vices}
\RCS$Revision: 440587 $
\RCS$HeadURL: svn+ssh://svn.cern.ch/reps/tdr2/papers/TOP-17-009/trunk/TOP-17-009.tex $
\RCS$Id: TOP-17-009.tex 440587 2018-01-06 18:49:03Z gzevi $
\newlength\cmsFigWidth
\ifthenelse{\boolean{cms@external}}{\setlength\cmsFigWidth{0.48\textwidth}}{\setlength\cmsFigWidth{0.9\textwidth}}
\ifthenelse{\boolean{cms@external}}{\providecommand{\cmsLeft}{upper\xspace}}{\providecommand{\cmsLeft}{left\xspace}}
\ifthenelse{\boolean{cms@external}}{\providecommand{\cmsRight}{lower\xspace}}{\providecommand{\cmsRight}{right\xspace}}

\newcommand{\sslumi}{35.9\fbinv}
\newcommand{\Njets}{\ensuremath{N_\text{jets}}\xspace}
\newcommand{\Nbjets}{\ensuremath{N_\text{b}}\xspace}
\newcommand{\Nleps}{\ensuremath{N_\ell}\xspace}
\newcommand{\tttt}{\ensuremath{\ttbar\ttbar}\xspace}
\newcommand{\ttZ}{\ensuremath{\ttbar\cPZ}\xspace}
\newcommand{\ttZgamma}{\ensuremath{\ttbar\cPZ/\gamma^{*}}\xspace}
\newcommand{\ttW}{\ensuremath{\ttbar\PW}\xspace}
\newcommand{\ttH}{\ensuremath{\ttbar\mbox{H}}\xspace}
\newcommand{\xsectttt}{\ensuremath{\sigma(\Pp\Pp\to\tttt})\xspace}
\newcommand{\POWHEGBOX} {{\textsc{powheg box}}\xspace}

\cmsNoteHeader{TOP-17-009}
\title{Search for standard model production of four top quarks with same-sign and multilepton final states in
 proton-proton collisions at $\sqrt{s} = 13\TeV$}

 \titlerunning{Search for four top quarks with same-sign and multilepton final states}

\date{\today}

\abstract{
A search for standard model production of four top quarks (\tttt) is reported using
events containing at least three leptons ($\Pe, \Pgm$) or a same-sign lepton pair.
The events are produced in proton-proton collisions at a center-of-mass energy of 13\TeV at the LHC, and the data sample, recorded in 2016, corresponds
to an integrated luminosity of 35.9\fbinv.
Jet multiplicity and flavor are used to enhance signal sensitivity,
and dedicated control regions are used to constrain the dominant backgrounds.
The observed and expected signal significances are, respectively, 1.6 and 1.0 standard deviations, and the \tttt cross section is measured to be
 $16.9^{+13.8}_{-11.4}$\unit{fb}, in agreement with  next-to-leading-order standard model predictions.  These results are also used to constrain the Yukawa coupling between the top quark and the Higgs boson to be less than 2.1 times its expected standard model value at 95\% confidence level.
}

\hypersetup{%
pdfauthor={CMS Collaboration},%
pdftitle={Search for standard model production of four top quarks with same-sign and multilepton final states in proton-proton collisions at sqrt(s) = 13 TeV},%
pdfsubject={CMS},%
pdfkeywords={CMS, physics, top quark}}

\maketitle
\section{Introduction}
\label{sec:intro}

\begin{figure}[!hbtp]
\centering
\includegraphics[width=.25\textwidth]{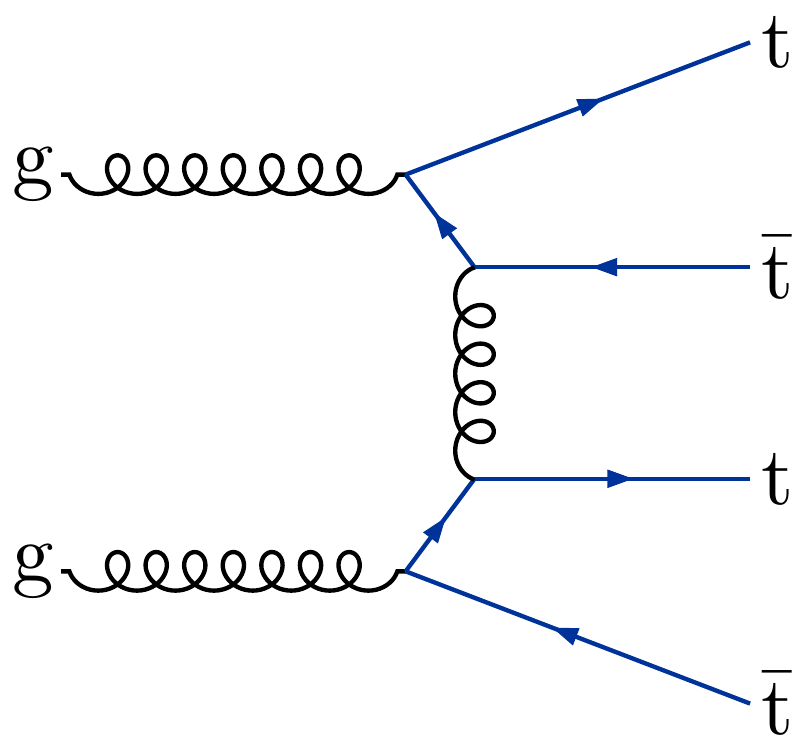}
\includegraphics[width=.25\textwidth]{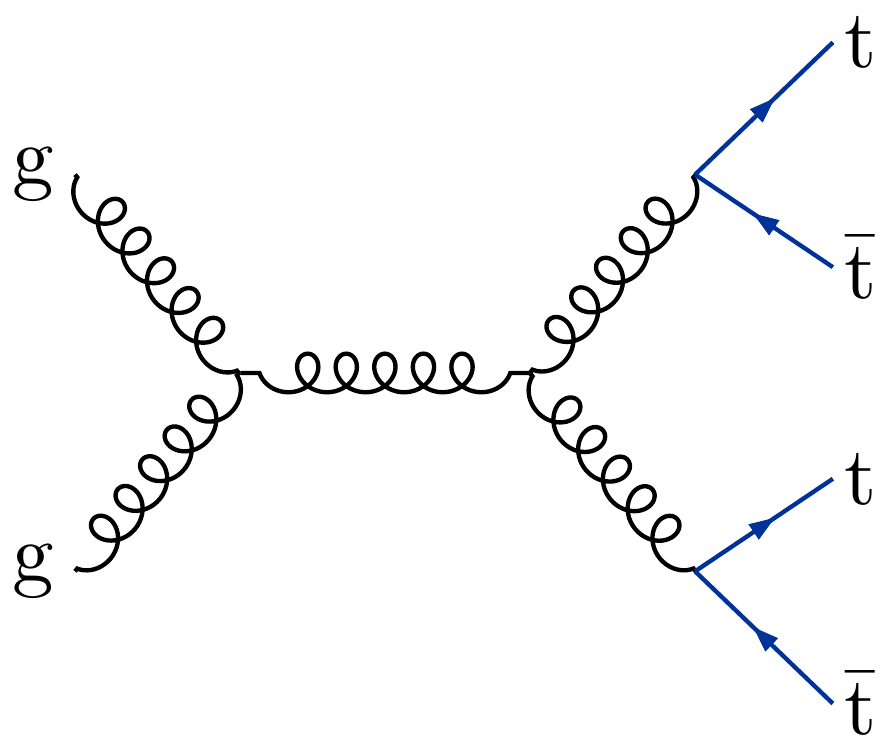}
\includegraphics[width=.25\textwidth]{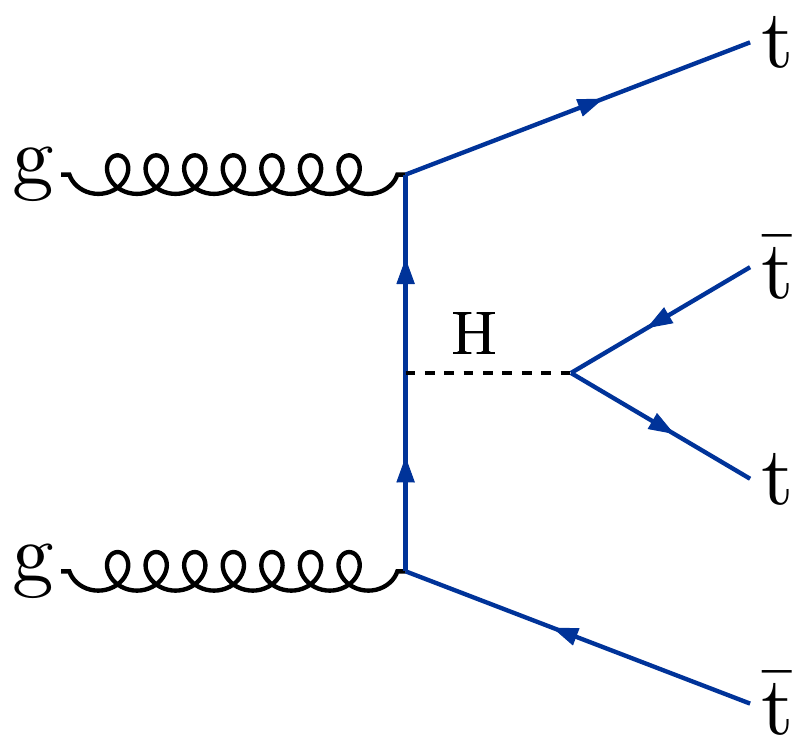}
\caption{Representative Feynman diagrams for \tttt production at LO in the SM.}
\label{fig:feyndiagrams}
\end{figure}

In the standard model (SM) the production of four top quarks (\tttt) is a rare process, with
representative leading-order (LO) Feynman diagrams shown in Fig.~\ref{fig:feyndiagrams}.
Many beyond-the-SM (BSM) theories predict an enhancement of the \tttt cross section, \xsectttt, such as
gluino pair production in the supersymmetry
framework~\cite{Ramond:1971gb,Golfand:1971iw,Neveu:1971rx,Volkov:1972jx,Wess:1973kz,Wess:1974tw,Fayet:1974pd,Nilles:1983ge,Martin:1997ns,Farrar:1978xj},
the pair production of scalar gluons~\cite{Plehn:2008ae,Calvet:2012rk},
and the production of a heavy pseudoscalar or scalar boson in association with a \ttbar pair in Type II two-Higgs-doublet models (2HDM)~\cite{Dicus:1994bm,Craig:2015jba,Craig:2016ygr}.
In addition, a top quark Yukawa coupling larger than expected in the SM can lead to a significant increase in \tttt production via an off-shell SM
Higgs boson~\cite{TopYukawaTTTT}.
The SM prediction for \xsectttt at $\sqrt{s} = 13 \TeV$ is $9.2^{+2.9}_{-2.4}$\unit{fb} at next-to-leading order (NLO)~\cite{MADGRAPH5}.
An alternative prediction of $12.2^{+5.0}_{-4.4}$\unit{fb} is reported in Ref. \cite{TopYukawaTTTT}, obtained from a LO calculation of $9.6^{+3.9}_{-3.5}$\unit{fb} and an
NLO/LO $K$-factor of 1.27 based on the $14\TeV$ calculation of Ref.~\cite{Bevilacqua2012}.

After the decays of the top quarks, the final state contains several jets resulting from the hadronization of light quarks and b quarks (b jets), and may contain isolated leptons and missing transverse momentum
depending on the decays of the \PW\ bosons~\cite{PDG}. Among these final states, the same-sign
 dilepton and the three- (or more) lepton final states, considering $\ell = \Pe, \mu$,
correspond to branching fractions in \tttt events of  8 and 1\%, respectively,
excluding the small contribution from $\PW\rightarrow\tau\nu$, which is included in selected events.
However, due to the low level of backgrounds, these channels are the most sensitive
to \tttt production in the regime with SM-like kinematic properties.
The ATLAS and CMS Collaborations at the CERN LHC have previously searched for SM
\tttt production in
$\sqrt{s}= 8$ and 13\TeV
\Pp\Pp~collisions~\cite{CMStttt8TeV,ATLAStttt8TeV,CMStttt2015,SS2016,ATLASssb13TeV}.
The most sensitive of these results is a re-interpretation of the CMS same-sign dilepton search
for BSM physics at $13~$\TeV~\cite{SS2016},
with an observed (expected) \tttt cross section upper limit (assuming no SM \tttt signal) of 42 ($27^{+13}_{-8}$)\unit{fb} at the 95\% confidence level (CL).

The previous search is inclusive, exploring the final
state with two same-sign leptons and at least two jets, using an integrated luminosity of \sslumi~\cite{SS2016}.
The analysis described in this  paper is based on the same data set and improves on the previous search
by optimizing the signal selection for sensitivity to SM \tttt production, by using an improved b jet identification algorithm,
and by employing background estimation techniques that
are adapted to take into account the higher jet and b jet multiplicity requirements in the signal regions.

\section{Background and signal simulation}
\label{sec:samples}

{\tolerance=800
Monte Carlo (MC) simulations at NLO are used to evaluate the \tttt signal acceptance and to estimate
the background from diboson (\PW\PZ, ZZ, $\PZ\Pgg$, $\PW^{\pm}\PW^{\pm}$) and triboson (\PW\PW\PW, \PW\PW\PZ, $\PW\PZ\PZ$, $\PZ\PZ\PZ$, $\PW\PW\Pgg$, $\PW\PZ\Pgg$) processes,
as well as from production of single top quarks ($\PQt\PZ\PQq$, $\PQt\Pgg$), or \ttbar produced in association with a boson (\ttW, \ttZgamma, \ttH).
These samples are generated using the NLO \MGvATNLO 2.2.2~\cite{MADGRAPH5} program with up to one additional parton in the matrix-element calculation,
except for $\PW^{\pm}\PW^{\pm}$ which is generated with up to two additional partons, and the  \PW\PZ,  ZZ and \ttH samples, which are generated with no additional partons with the \POWHEGBOX~v2~\cite{Melia:2011tj,Nason:2013ydw} program.
The \ttZgamma\ sample with $\PZ/\gamma^{*}\rightarrow\ell\ell$ is generated with a dilepton invariant mass greater than $1\GeV$.
The LO \MGvATNLO generator, scaled to NLO cross sections,
is used to estimate the $\PW\Pgg$ and $\ttbar \Pgg$ processes with up to three additional partons.
Other rare backgrounds, such as \ttbar production in association with dibosons ($\ttbar\PW\PW$, $\ttbar\PW\PZ$, $\ttbar\PZ\PZ$, $\ttbar\PW\PH$,
$\ttbar\PZ\PW$, $\ttbar\PH\PH$), triple top quark production
($\ttbar\PQt$, $\ttbar\PQt\PW$), and $\PQt\PW\PZ$ are generated using LO \MGvATNLO without additional partons, and scaled to NLO cross sections~\cite{LHCHXSWG4}.
The NNPDF3.0LO (NNPDF3.0NLO)~\cite{Ball:2014uwa} parton distribution
functions (PDFs) are used to generate all LO (NLO) samples.
Parton showering and hadronization,  as well as $\PW^{\pm}\PW^{\pm}$ from double-parton-scattering, are modeled by the
\PYTHIA~8.205~\cite{Sjostrand:2007gs} program, while
the MLM~\cite{Alwall:2007fs} and FxFx~\cite{Frederix:2012ps} prescriptions are employed in matching
additional partons in the matrix-element calculations to parton showers in the LO and NLO samples, respectively.
The top quark mass in the generators is set to $172.5\GeV$.  The \GEANTfour package~\cite{Geant} is used to model the response of the CMS detector.
Additional proton-proton interactions (pileup) within the same or nearby bunch crossings are also included in the simulated events.
\par}

To improve the MC modeling of the
multiplicity of additional jets from initial-state radiation (ISR),
simulated \ttW and \ttZgamma events are reweighted based on the number of ISR jets ($N_{\text{jets}}^\mathrm{ISR}$).
The reweighting is based on a comparison of the light-flavor jet
multiplicity in dilepton \ttbar events in data and simulation.
The method requires exactly two jets identified as originating from b quarks in dilepton \ttbar events, and assumes that all other jets are
from ISR.
Weighting factors are obtained as a function of $N_{\text{jets}}^\mathrm{ISR}$ to bring data and MC into agreement.
These weights are then applied, keeping the total cross section constant, to \ttW and \ttZgamma MC as a function of the number of jets not originating from top quark, \PW, or \PZ decays.
To improve the modeling of the flavor of additional jets, the simulation
is also corrected to account for the measured ratio of $\ttbar \bbbar / \ttbar \text{jj}$ cross sections
reported in Ref.~\cite{CMSttbb}.
More details on these corrections and their uncertainties are provided in Section~\ref{sec:systematics}.

\section{ The CMS detector and event reconstruction}
\label{sec:cms}

The central feature of the CMS detector is a superconducting solenoid of 6\unit{m} internal diameter, providing a magnetic field of 3.8\unit{T}. Within the solenoid volume are a silicon pixel and strip tracker, a lead tungstate crystal electromagnetic calorimeter (ECAL), and a brass and scintillator hadron calorimeter (HCAL), each composed of a barrel and two endcap sections. Forward calorimeters extend the pseudorapidity ($\eta$) coverage provided by the barrel and endcap detectors. Muons are measured in gas-ionization detectors embedded in the steel flux-return yoke outside the solenoid.
A more detailed description of the CMS detector, together with a definition of the coordinate system used and the relevant kinematic variables, can be found in Ref.~\cite{Chatrchyan:2008zzk}.

Events of interest are selected using a two-tiered trigger system~\cite{Khachatryan:2016bia}. The first level (L1), composed of custom hardware processors, uses information from the calorimeters and muon detectors to select events at a rate of around 100\unit{kHz} within a time interval of less than 4\mus. The second level, known as the high-level trigger (HLT), consists of a farm of processors running a version of the full event reconstruction software optimized for fast processing, and reduces the event rate to less than 1\unit{kHz} before data storage.

Events are processed using the particle-flow (PF) algorithm~\cite{CMS-PRF-14-001}, which reconstructs and identifies each individual particle with an optimized combination of information from the various elements of the CMS detector. The energy of photons is directly obtained from the ECAL measurement. The energy of electrons is determined from a combination of the electron momentum at the primary interaction vertex as determined by the tracker, the energy of the corresponding ECAL cluster, and the energy sum of all bremsstrahlung photons spatially compatible with the electron track~\cite{Khachatryan:2015hwa}. The momentum of muons is obtained from the curvature of the corresponding track, combining information from the silicon tracker and the muon system~\cite{Chatrchyan:2012xi}. The energy of charged hadrons is determined from a combination of their momentum measured in the tracker and the matching ECAL and HCAL energy deposits, corrected
for the response function of the calorimeters to hadronic showers. The energy of neutral hadrons is obtained from the corresponding corrected ECAL and HCAL energies.

Hadronic jets are clustered from neutral PF candidates and charged PF candidates associated with the primary vertex, using the anti-\kt algorithm~\cite{Cacciari:2008gp, Cacciari:2011ma} with a distance parameter of 0.4.
The jet momentum is determined as the vectorial sum of all PF candidate momenta in the jet. An offset correction is applied to jet energies to take into account the contribution from pileup.
Jet energy corrections are derived from simulation, and are improved with in situ measurements of the energy balance in dijet, multijet, \Pgg+jet and leptonically decaying Z+jet events~\cite{Chatrchyan:2011ds,Khachatryan:2016kdb}. Additional selection criteria are applied to each event to remove spurious jet-like features originating from isolated noise patterns in certain HCAL regions.
Jets originating from \cPqb\ quarks are identified as b-tagged jets using a deep neural network algorithm~\cite{CMS-DP-2017-005},
with a working point chosen such that the efficiency to identify a b jet is  55--70\% for a jet transverse momentum (\pt) between 20 and 400\GeV.
The misidentification rate for a light-flavor jet is 1--2\% in the same jet \pt range.
The vector \ptvecmiss is defined as the projection on the plane perpendicular to the beams of the negative vector sum of the momenta of all reconstructed PF candidates in an event~\cite{CMS-PAS-JME-16-004}. Its magnitude, called missing transverse momentum, is referred to as \ptmiss. The scalar \pt sum of all jets in an event is referred to as \HT.

\section{Event selection and search strategy}
\label{sec:selection}

The definitions of objects and the baseline event selection follow closely those of Refs.~\cite{SS2016,SUS-15-008}.
Electron identification is based on a multivariate discriminant using shower shape and track quality variables,
while for muons it is based on the quality of the geometrical matching between the tracker and muon system measurements.
Isolation and impact parameter requirements are applied to both lepton flavors, as well as specific selections designed
to improve the accuracy of the charge reconstruction.
The combined reconstruction and identification efficiency is in the range of 45--70\% (70--90\%)  for electrons (muons),
increasing as a function of \pt and converging to the maximum value for $\pt>$ 60\GeV.
The number of leptons (\Nleps), the number of jets (\Njets),
and the number of b-tagged jets (\Nbjets) are counted after the application of the basic
kinematic requirements summarized in Table~\ref{tab:objetselection}.

\begin{table}[!hbtp]
\centering
\topcaption{Kinematic requirements for leptons and jets.}
\label{tab:objetselection}
\begin{tabular}{c|cc}
Object        & \pt\,(\GeVns{})            & $\abs{\eta}$ \\
\hline
Electrons     & $>$20 & $<$2.5 \\
Muons         & $>$20 & $<$2.4 \\
Jets          & $>$40 & $<$2.4 \\
b-tagged jets & $>$25 & $<$2.4 \\
\end{tabular}
\end{table}

Signal events are selected using triggers that require two leptons with $\pt > 8\GeV$ and $\HT > 300\GeV$.
The trigger efficiency is greater than 95\% for di-electron ($\Pe\Pe$) and electron-muon ($\Pe\Pgm$) events and about 92\% for di-muon ($\Pgm\Pgm$) events.
The baseline selections require $\HT > 300\GeV$ and $\ptmiss>50\GeV$,
at least two jets ($\Njets \geq 2$), at least two b-tagged jets ($\Nbjets \geq 2$), a leading lepton with $\pt > 25\GeV$,
and a second lepton of the same charge with $\pt > 20\GeV$.
To reduce the background from Drell--Yan with a charge-misidentified electron, events with same-sign electron pairs with mass below 12\GeV are rejected.
Events where a third lepton with \pt larger than 5 (7)\GeV for muons (electrons) forms an opposite-sign (OS) same-flavor
pair with mass below 12\GeV or between 76 and 106\GeV are also rejected.
If the third lepton has $\pt > 20\GeV$ and the invariant mass of the pair is between 76 and 106\GeV, these rejected events
 are used to populate a \ttZ background control region (CRZ).
The signal acceptance in the baseline region, including the leptonic \PW\ boson branching fraction, is approximately 1.5\%.
After these requirements, we define eight mutually exclusive signal regions (SRs) and a control region  for the \ttW background (CRW),  based on \Njets, \Nbjets, and \Nleps,
as detailed in Table~\ref{tab:SRDef}.
The observed and predicted yields in the control and signal regions are used to measure \xsectttt, following the procedure described in Sec.~\ref{sec:results}.

\begin{table}[ht]
\centering
\topcaption{\label{tab:SRDef} Definitions of the eight SRs and the two control regions for \ttW (CRW) and \ttZ (CRZ).}
\begin{tabular}{c|c|c|c}
$\Nleps$ &  $\Nbjets$ & $\Njets$ & Region \\ \hline
\multirow{7}{*}{2} & \multirow{4}{*}{2} & $\leq$5 &  CRW  \\ \cline{3-4}
                   &                    &    6     &  SR1  \\ \cline{3-4}
                   &                    &    7     &  SR2  \\ \cline{3-4}
                   &                    & $\geq$8 &  SR3  \\ \cline{2-4}
                   & \multirow{2}{*}{3} &  5, 6    &  SR4  \\ \cline{3-4}
                   &                    & $\geq$7 &  SR5  \\ \cline{2-4}
                   &   $\geq$4         & $\geq$5 &  SR6  \\ \hline
\multirow{2}{*}{$\geq$3} & 2                  & $\geq$5 &  SR7  \\ \cline{2-4}
                   &   $\geq$3         & $\geq$4 &  SR8  \\ \hline
\multicolumn{3}{c|}{Inverted Z veto} & CRZ  \\
\end{tabular}
\end{table}

\section{Backgrounds}
\label{sec:backgrounds}
The main backgrounds to the \tttt process in the same-sign dilepton and three- (or more) lepton final states
 arise from rare multilepton processes, such as \ttW, \ttZgamma, and \ttH ($\mathrm{H} \to \mathrm{WW}$),
and single-lepton or OS dilepton processes with an additional ``nonprompt lepton''.
Nonprompt leptons consist of electrons from conversions of photons in jets and leptons from the decays of heavy- or light-flavor hadrons.
In this category we include also hadrons misidentified as leptons.
The minor background from OS dilepton events with a charge-misidentified lepton is also taken into account.

Rare multilepton processes are estimated using simulated events.
Control regions are used to constrain the normalization of the \ttW and \ttZ backgrounds,
as described in Section~\ref{sec:results}, while for other processes the normalization is based on the NLO cross sections
referenced in Section~\ref{sec:samples}.
Processes such as the associated production of a \ttbar pair with a pair of bosons ($\PW$, $\PZ$,  \PH) are grouped into a ``$\ttbar\mathrm{VV}$'' category.
Associated photon production processes such as $\PW\Pgg$, $\PZ\Pgg$, $\ttbar \Pgg$, and $\PQt\Pgg$,
where an electron is produced in an unidentified photon conversion, are grouped into a ``X\Pgg'' category.
All residual processes with very small contributions, including diboson (\PW\PZ, ZZ, $\PW^{\pm}\PW^{\pm}$ from single- and double-parton scattering),
triboson (\PW\PW\PW, \PW\PW\PZ, $\PW\PZ\PZ$, $\PZ\PZ\PZ$, $\PW\PW\Pgg$, $\PW\PZ\Pgg$), and rare single top quark ($\PQt\PZ\PQq$, $\PQt\PW\PZ$) and triple top quark processes
($\ttbar\PQt$ and $\ttbar\PQt\PW$), are grouped into a ``Rare'' category.

The nonprompt lepton and charge-misidentified lepton backgrounds are estimated following the methods described in Ref.~\cite{SS2016}.
For nonprompt leptons, an estimate referred to as the ``tight-to-loose'' method defines two control regions
by modifying the  lepton identification (including isolation) and event kinematic requirements, respectively.
An ``application region'' is defined for every SR by requiring at least one lepton to fail the standard identification (``tight'')
while satisfying a more relaxed one (``loose'').
To obtain the nonprompt lepton background estimate in the corresponding SR, the event yield in each application region is weighted
by a factor of $\epsilon_\mathrm{TL} / (1-\epsilon_\mathrm{TL})$ for each lepton failing the tight requirement.
The $\epsilon_\mathrm{TL}$ parameter is the probability that a nonprompt lepton that satisfies a loose lepton selection also satisfies the tight selection.
It is extracted as a function of lepton flavor and kinematic properties from a ``measurement region'' that consists of
a single-lepton events with event kinematic properties designed to suppress the $\PW \to \ell \nu$ contribution.

For charge-misidentified leptons, an OS dilepton control region is defined for each same-sign dilepton signal region.
Its yield is then weighted by the charge misidentification probability estimated in simulation, which ranges between
$10^{-5}$ and $10^{-3}$ for electrons and  is  negligible for muons.

\section{Systematic uncertainties}
\label{sec:systematics}
The sources of experimental and theoretical uncertainty for the data and simulations
are summarized in Table~\ref{tab:systSummary}.
The uncertainty in the integrated luminosity is 2.5\%~\cite{cmsLumi2016}.
The simulation is reweighted to match the distribution in the number of pileup
collisions per event in data. The uncertainty in the inelastic cross section  propagated to the final yields
provides an uncertainty of at most 6\%.

Trigger efficiencies are measured with an uncertainty of 2\% in an independent data sample selected using single-lepton triggers.
Lepton-efficiency scale factors, used to account for differences in the reconstruction and identification efficiencies between data and simulation,
are measured using a ``tag-and-probe'' method in data enriched in $\PZ\to\ell\ell$ events~\cite{Khachatryan:2015hwa,Chatrchyan:2012xi}.
The scale factors are applied to all simulated processes with an uncertainty per lepton of approximately 3\% for muons and 4\% for electrons.

The uncertainty in the calibration of the jet energy scale depends on the \pt and $\eta$ of the jet
and results in a 1--15\% variation in the event yield in a given SR.
The uncertainty due to the jet energy resolution is estimated by broadening the
resolution in simulation~\cite{Khachatryan:2016kdb},
and the resulting
effect is a change of 1--5\% in the SR yields.
The b tagging efficiency in simulation is corrected using scale factors determined from efficiencies measured in data and simulation~\cite{CMS-PAS-BTV-15-001}.
The uncertainty in the measured scale factors results in an overall effect between 1 and 15\%, again depending on the SR.

As mentioned in Section~\ref{sec:samples}, \ttW and \ttZgamma simulated events are reweighted to match the number of additional jets observed in data.
The reweighting factors vary between 0.92 for $N_{\text{jets}}^\mathrm{ISR}=1$ and 0.77 for $N_{\text{jets}}^\mathrm{ISR} \geq 4$.
Half of the difference from unity is taken as a systematic uncertainty in these reweighting factors to cover differences observed between data and simulation
when the factors are used to reweight simulation in a control sample enriched in single-lepton \ttbar events.
Uncertainties in the reweighting factors are treated as correlated among regions.
Simulated \ttW and \ttZgamma events with two b quarks not originating from top
quark decay are also weighted to account for the CMS measurement of the ratio of cross sections
$\sigma({\ttbar\bbbar})/\sigma({\ttbar\mathrm{jj}})$, which was found to be a factor of $1.7\pm0.6$ larger than the MC prediction~\cite{CMSttbb}.
In signal regions requiring four b-tagged jets, where the effect is
dominant, this results in a systematic uncertainty of up to 15\% on the total background prediction.
In signal regions requiring three b-tagged jets, the dominant origin of the additional b-tagged jet is a charm quark from a \PW\ decay, so the effect is negligible.

Uncertainties in the renormalization and factorization scales (varied by a factor of two) and from the choice of PDF~\cite{Botje:2011sn,Alekhin:2011sk} affect
the number of events expected (normalization) in the simulated background
processes, as well as the acceptance for the \tttt signal.
The effects of these uncertainties on the relative distribution of events
in the signal regions (shape)
are also considered.
For the \ttW and \ttZgamma backgrounds, the normalization uncertainty is 40\%,
while for \ttH a 50\% normalization uncertainty reflects the signal strength of $1.5 \pm 0.5$ measured by CMS~\cite{CMS-PAS-HIG-17-004}.
The processes in the Rare category along with X\Pgg\ and $\ttbar\mathrm{VV}$, many of which have never been observed, are expected to give small contributions
to the event yields in the signal regions.  We assign separate 50\% normalization uncertainties to each of these three categories.
The shape uncertainty resulting from variations of the renormalization and factorization scales is as large as 15\% for the \ttW, \ttZgamma, and \ttH backgrounds, and 10\% for the \tttt signal,
while the effect from the PDF is only 1\%.
For the signal, the uncertainty in the acceptance from variations of the scales (PDFs) is 2\%\,(1\%).
In addition, for the \tttt signal, the scales that determine ISR and final-state radiation (FSR) in the parton shower are also varied,
resulting in a 6\% change in the acceptance and shape variations as large as 15\%.

For nonprompt and charge-misidentified lepton backgrounds, the statistical uncertainty
from the application region depends on the SR considered.
The background from misidentified charge is assigned a systematic uncertainty of 20\%, based on comparisons of the expected number of same-sign events estimated from an
OS control sample and the observed same-sign yield in a control sample enriched in $\PZ\to\Pep\Pem$ events with one electron or positron
having a misidentified charge.

In addition to the statistical uncertainty, the nonprompt lepton background is assigned
an overall normalization uncertainty of 30\% to cover variations observed in closure tests performed with simulated multijet and \ttbar events.
This uncertainty is increased to 60\% for electrons with $\pt > 50\GeV$, to account for trends observed at high \pt in the closure tests.
We also include an uncertainty related to the subtraction of events with prompt leptons (from electroweak processes with a \PW{} or \PZ{} boson) in the measurement region, which has an effect between 1\% and 50\%,
depending on the SR.
The prompt lepton contamination was also checked in the application region, where it was found to be below 1\%.

Experimental uncertainties are treated as correlated among signal regions for all signal and background processes.
Systematic uncertainties in data-driven estimates and theoretical uncertainties are treated as uncorrelated between processes, but correlated among signal regions.
Statistical uncertainties from the limited number of  simulated events or in the number of events in data control regions are considered  uncorrelated.

\begin{table}[!hbtp]
\centering
\topcaption{
   Summary of the sources of uncertainty and their effect on signal and background yields.
    The first group lists experimental and theoretical uncertainties
    in simulated signal and background processes.
    The second group lists normalization uncertainties in the estimated backgrounds.
    }
    \label{tab:systSummary}
\begin{tabular}{l|c}
Source                                      & Uncertainty (\%) \\
\hline
Integrated luminosity                       & 2.5 \\
Pileup                                      & 0--6 \\
Trigger efficiency                          & 2 \\
Lepton selection                            & 4--10 \\
Jet energy scale                            & 1--15 \\
Jet energy resolution                       & 1--5 \\
$\cPqb$ tagging                             & 1--15 \\
Size of simulated sample               & 1--10 \\
Scale and PDF variations                    & 10--15 \\
ISR/FSR (signal)                              & 5--15 \\
\hline
$\ttbar \PH$ (normalization)                & 50 \\
Rare, X\Pgg,  $\ttbar\mathrm{VV}$ (norm.) & 50 \\
$\ttZgamma$, $\ttbar \PW$  (normalization) & 40 \\
Charge misidentification                    & 20 \\
Nonprompt leptons                           & 30--60 \\
\end{tabular}
\end{table}

\section{Results and interpretation}
\label{sec:results}

\begin{figure*}[hbtp]
\centering
\includegraphics[width=.45\textwidth]{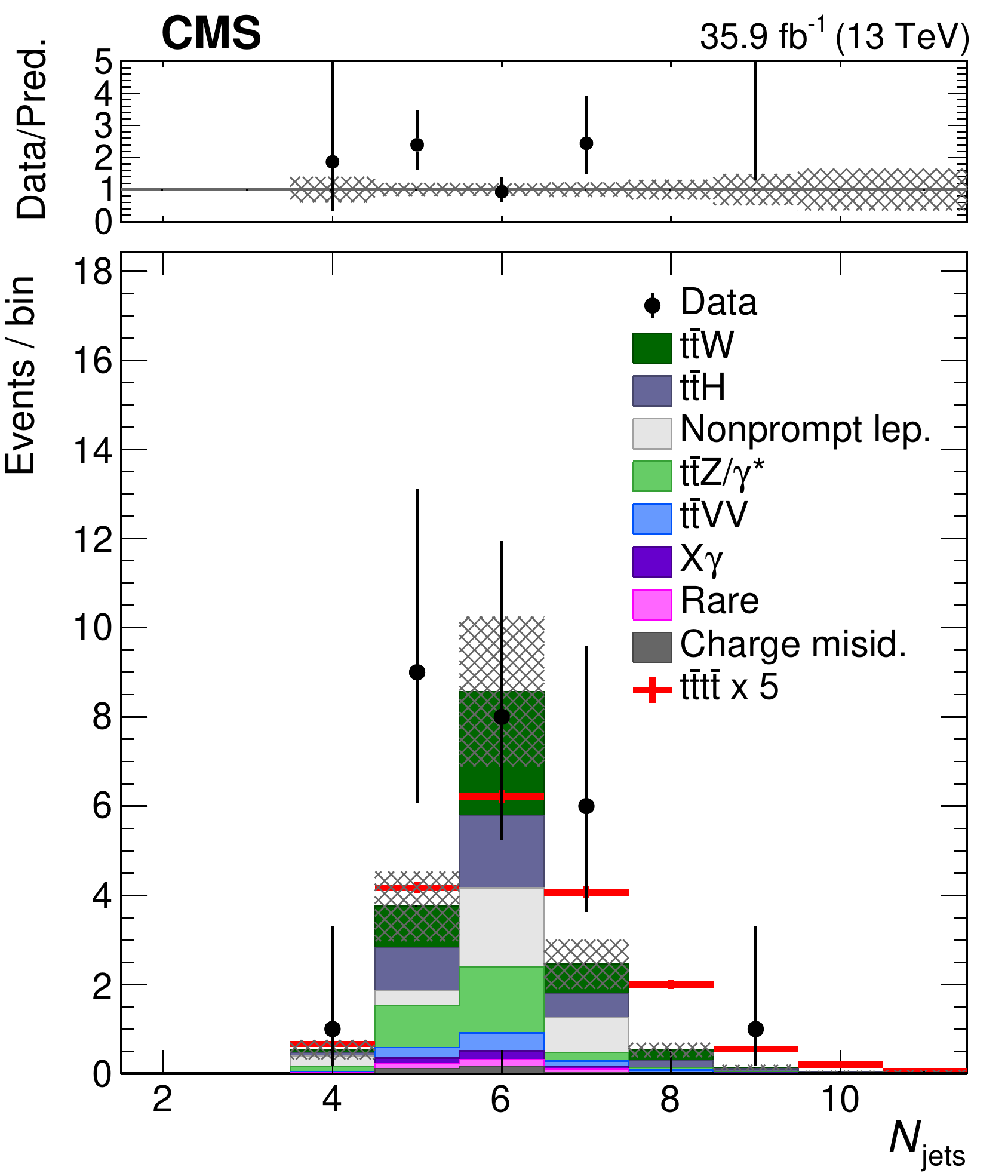}
\includegraphics[width=.45\textwidth]{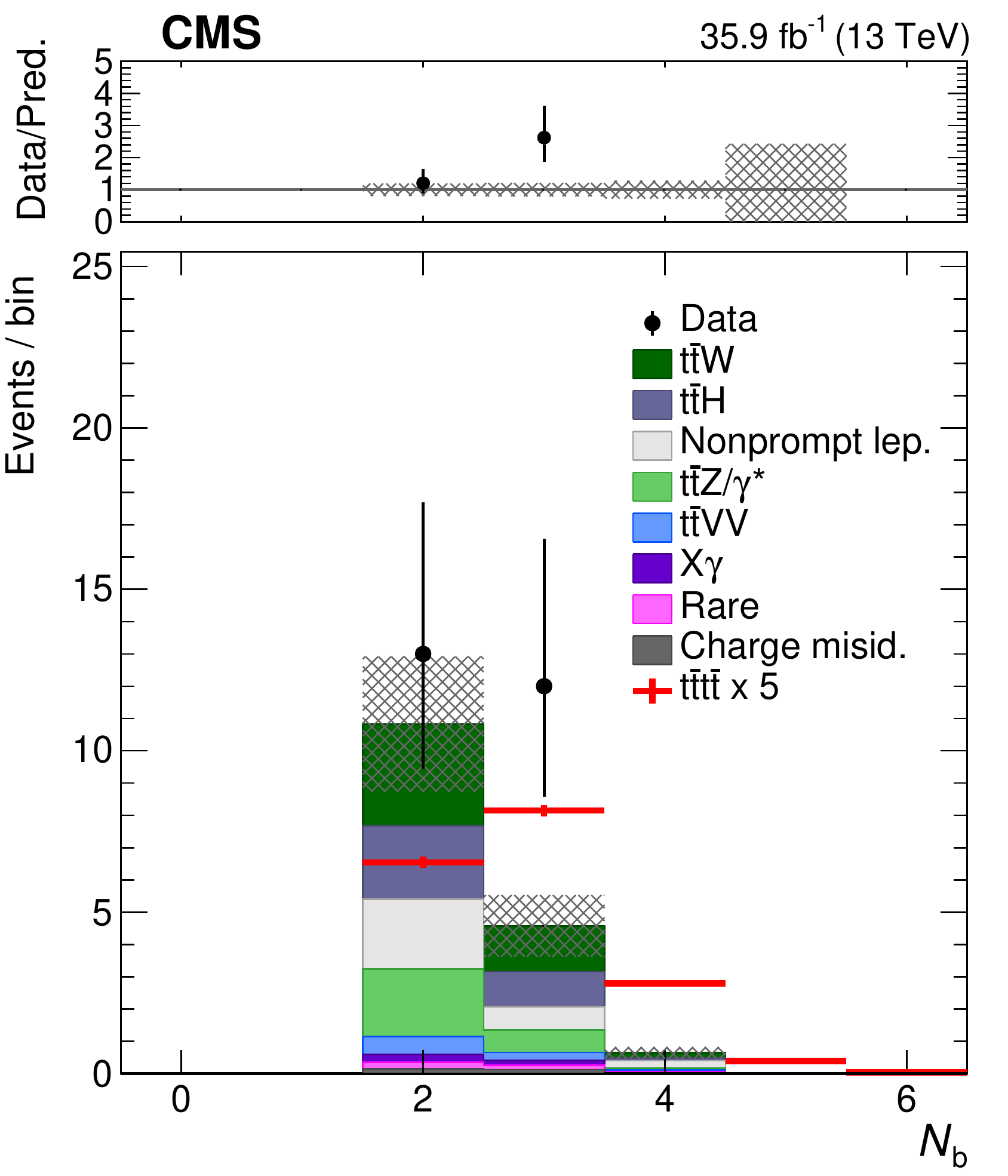}
\includegraphics[width=.45\textwidth]{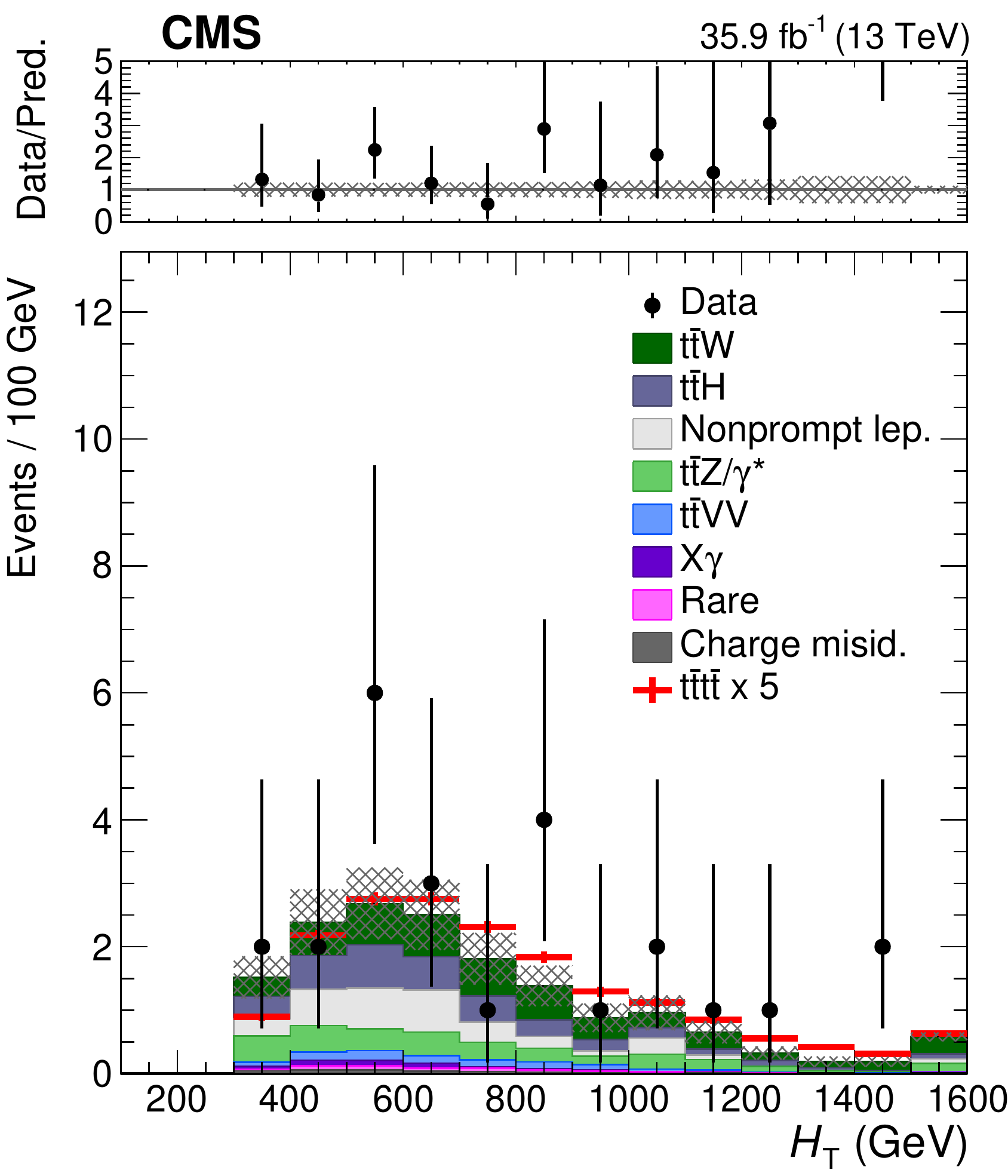}
\includegraphics[width=.45\textwidth]{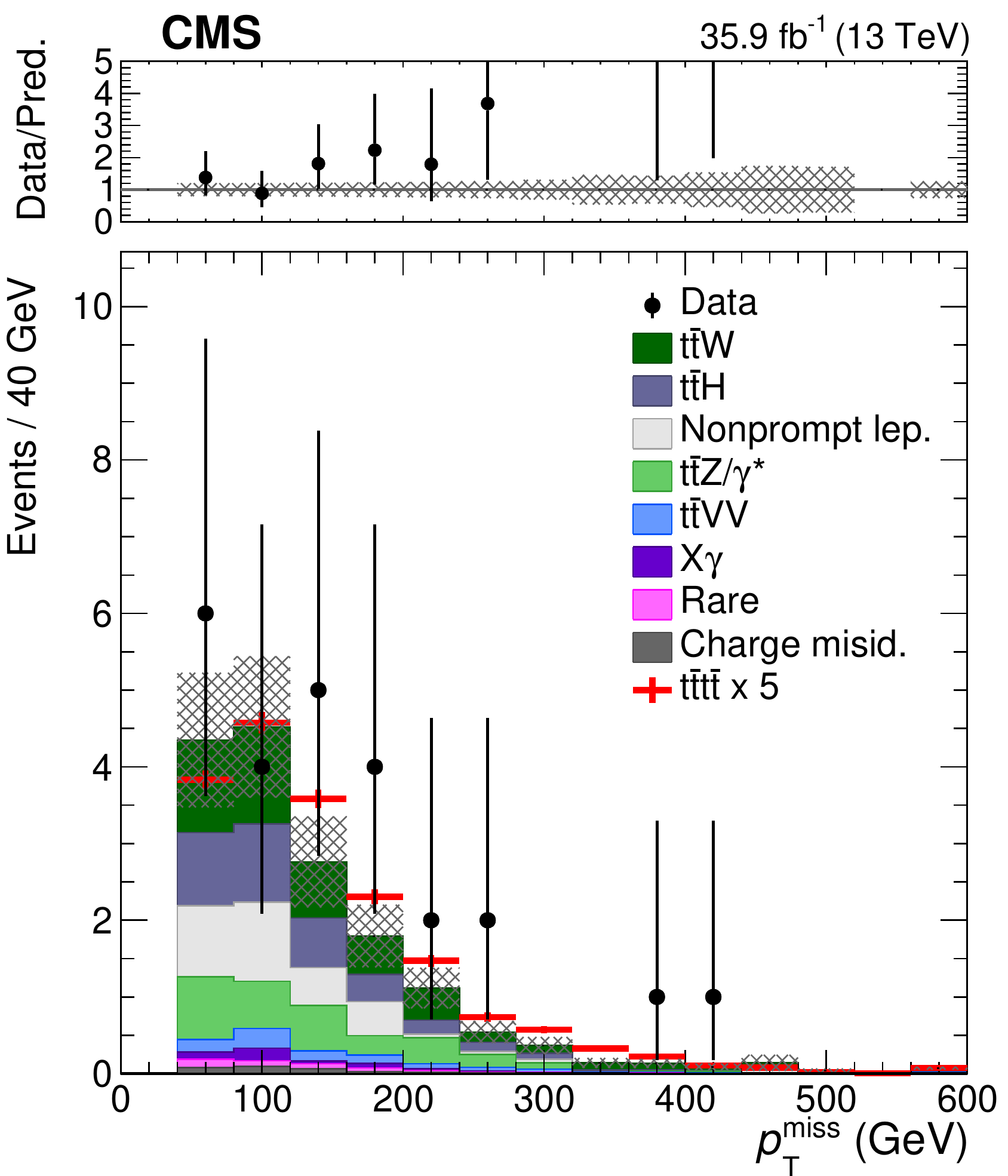}
\caption{Distributions in \Njets (upper left), \Nbjets (upper right), \HT (lower left), and \ptmiss (lower right) in the signal regions (SR~1--8), before fitting to data,
where the last bins include the overflows. The hatched areas represent the total uncertainties in the SM background predictions,
while the solid lines represent the \tttt signal, scaled up by a factor of 5,
assuming the SM cross section from Ref.~\cite{MADGRAPH5}.
 The upper panels show the ratios of the observed event yield to the total background prediction.
Bins without a data point have no observed events.
}
\label{fig:kinemsr}
\end{figure*}

\begin{figure*}[hbtp]
\centering
\includegraphics[width=.45\textwidth]{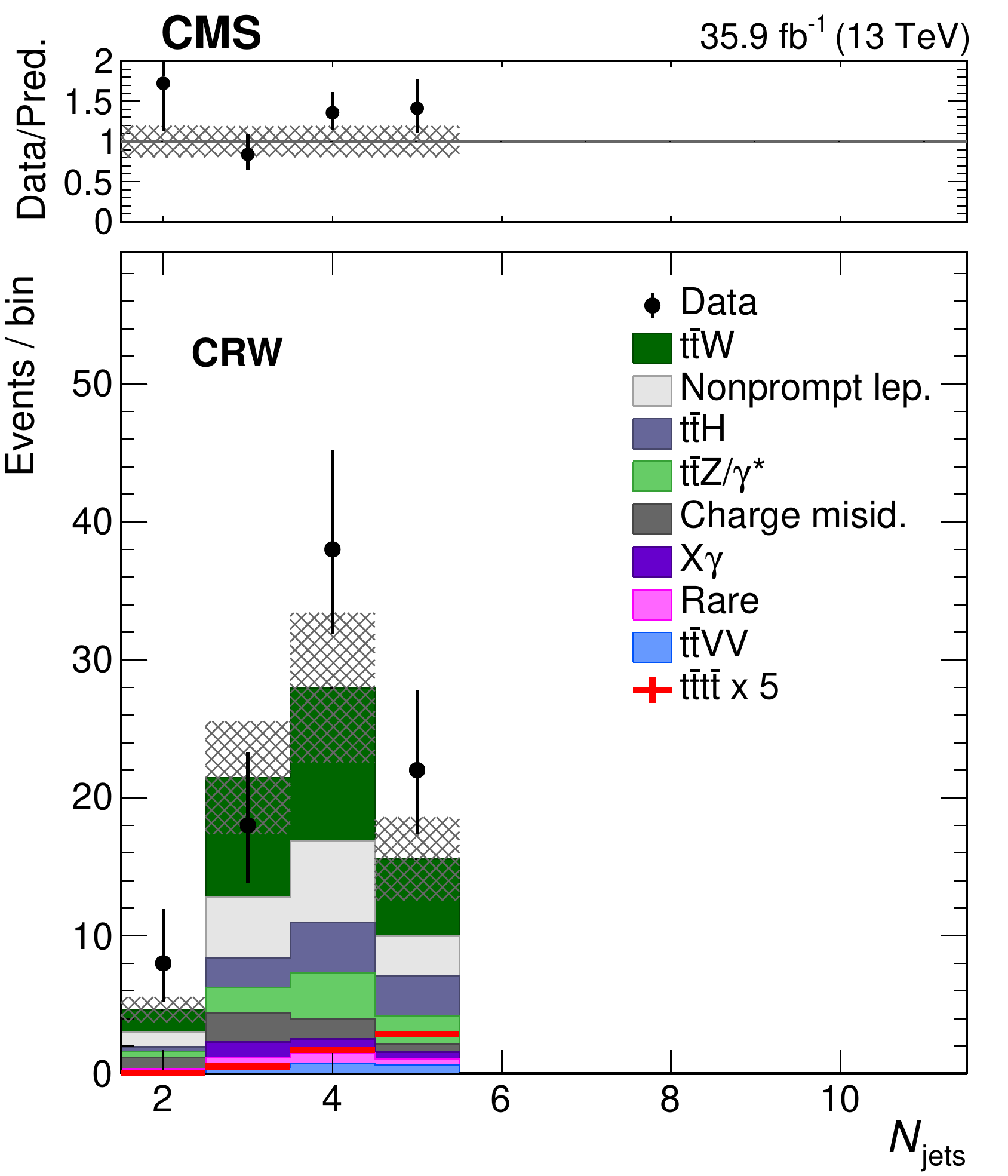}
\includegraphics[width=.45\textwidth]{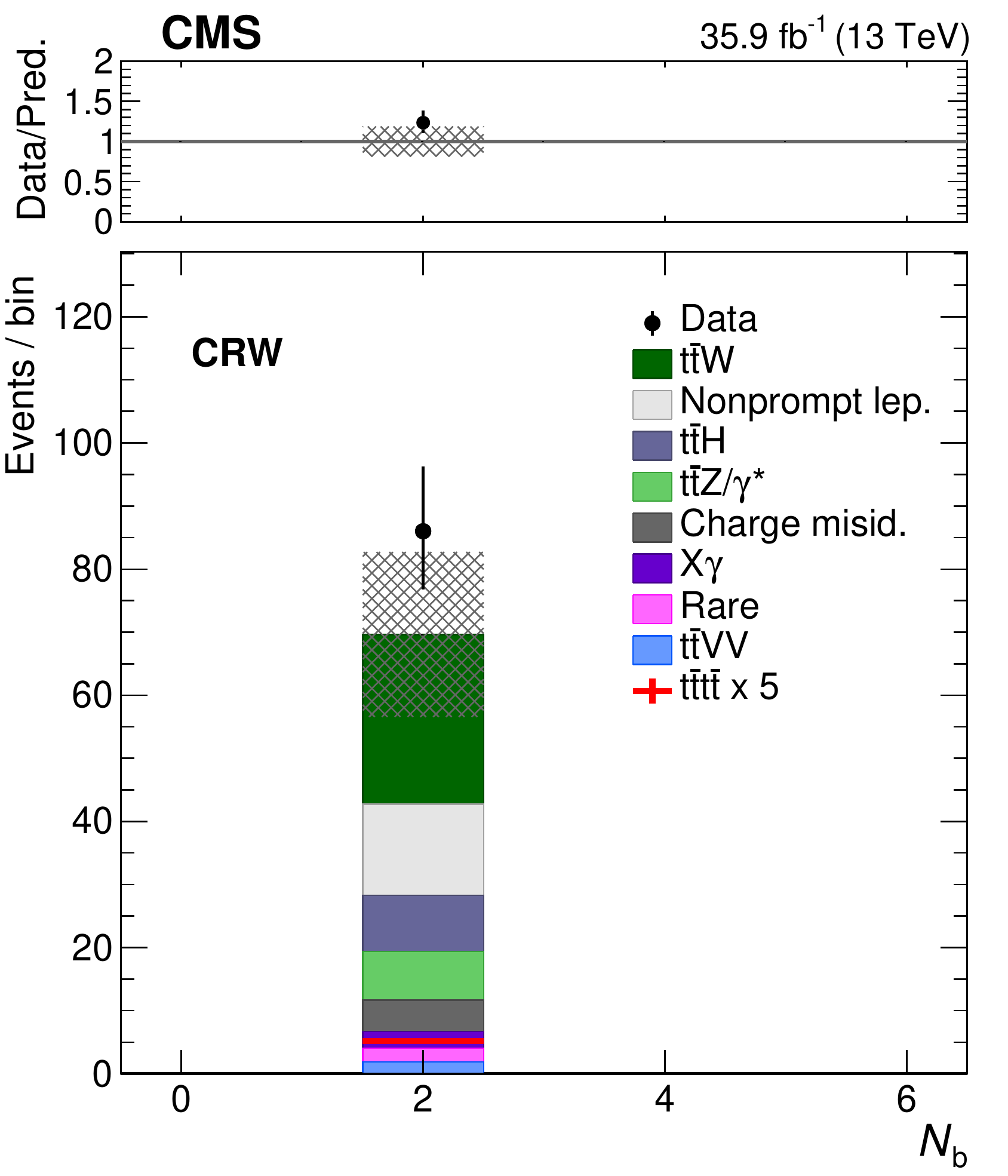}
\includegraphics[width=.45\textwidth]{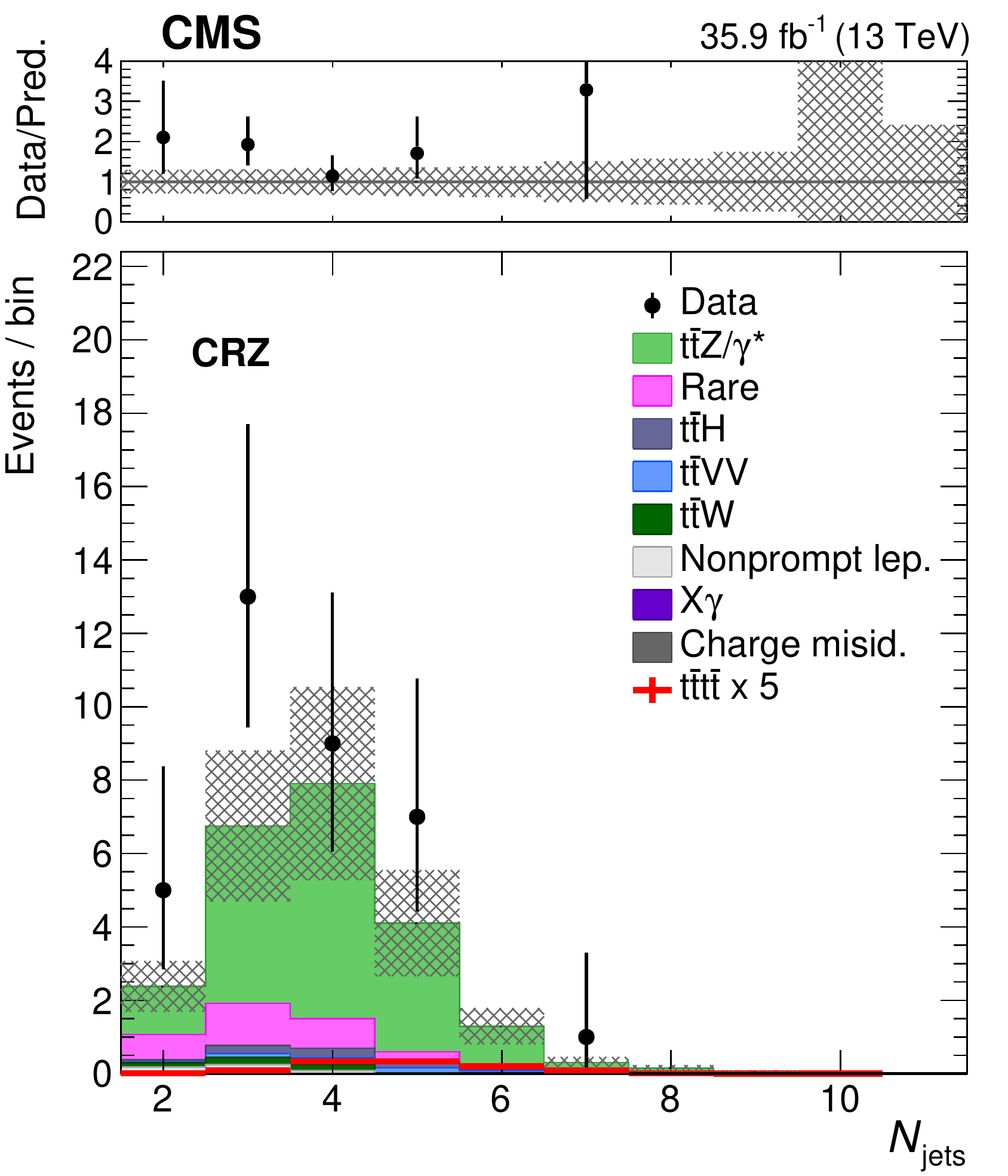}
\includegraphics[width=.45\textwidth]{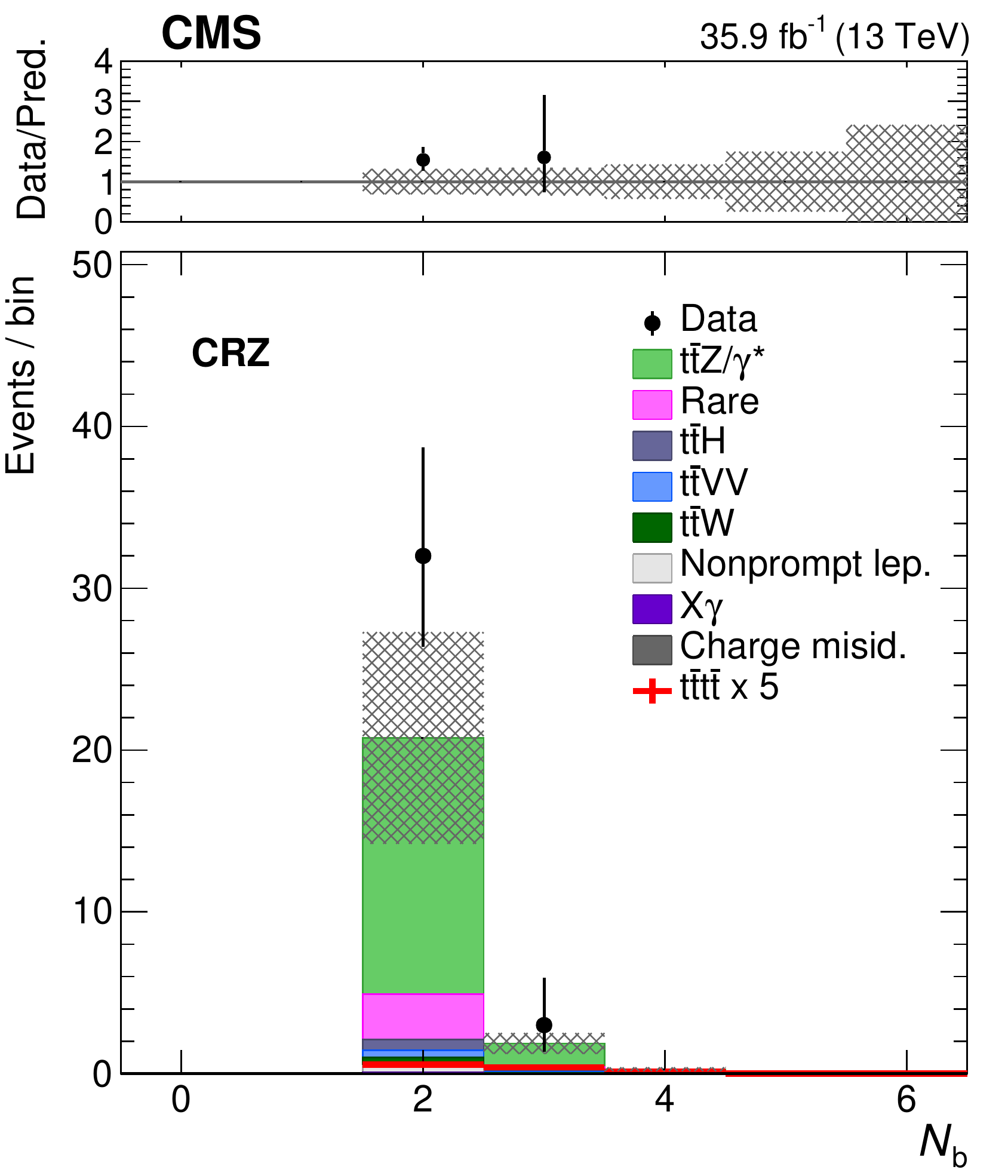}
\caption{Distributions in \Njets and \Nbjets in \ttW (upper) and \ttZ (lower) control regions, before fitting to data.
The hatched area represents the  uncertainty in the
SM background prediction, while the solid line represents the \tttt signal, scaled up by a factor of 5,
assuming the SM cross section from Ref.~\cite{MADGRAPH5}.
 The upper panels show the ratios of the observed event yield to the total background prediction.
Bins without a data point have no observed events.
}
\label{fig:kinemcr}
\end{figure*}

The  properties of events in the signal regions (SR~1--8 as defined in Table~\ref{tab:SRDef}) are shown in
Fig.~\ref{fig:kinemsr}, where distributions of the main kinematic variables in the data
(\Njets, \Nbjets, \HT, and \ptmiss) are compared to
SM background predictions.
The \Njets and \Nbjets distributions for CRW and CRZ are shown in Fig.~\ref{fig:kinemcr}.
In both figures we overlay the expected SM \tttt signal, scaled by a factor of 5.
The SM predictions are generally consistent with the observations,
with some possible underestimation in CRW and CRZ.

\begin{figure}[hbtp]
\centering
\includegraphics[width=.45\textwidth]{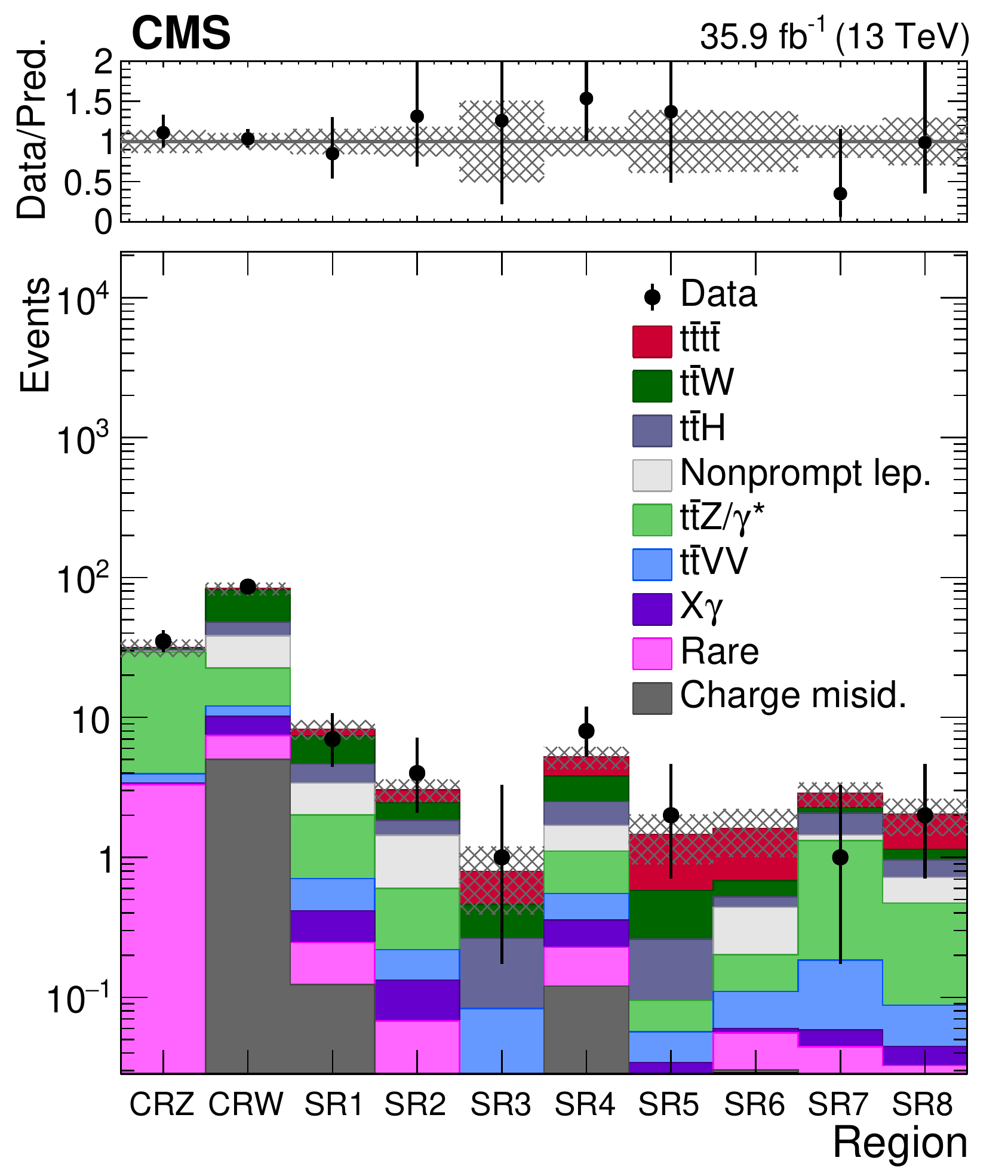}
\includegraphics[width=.45\textwidth]{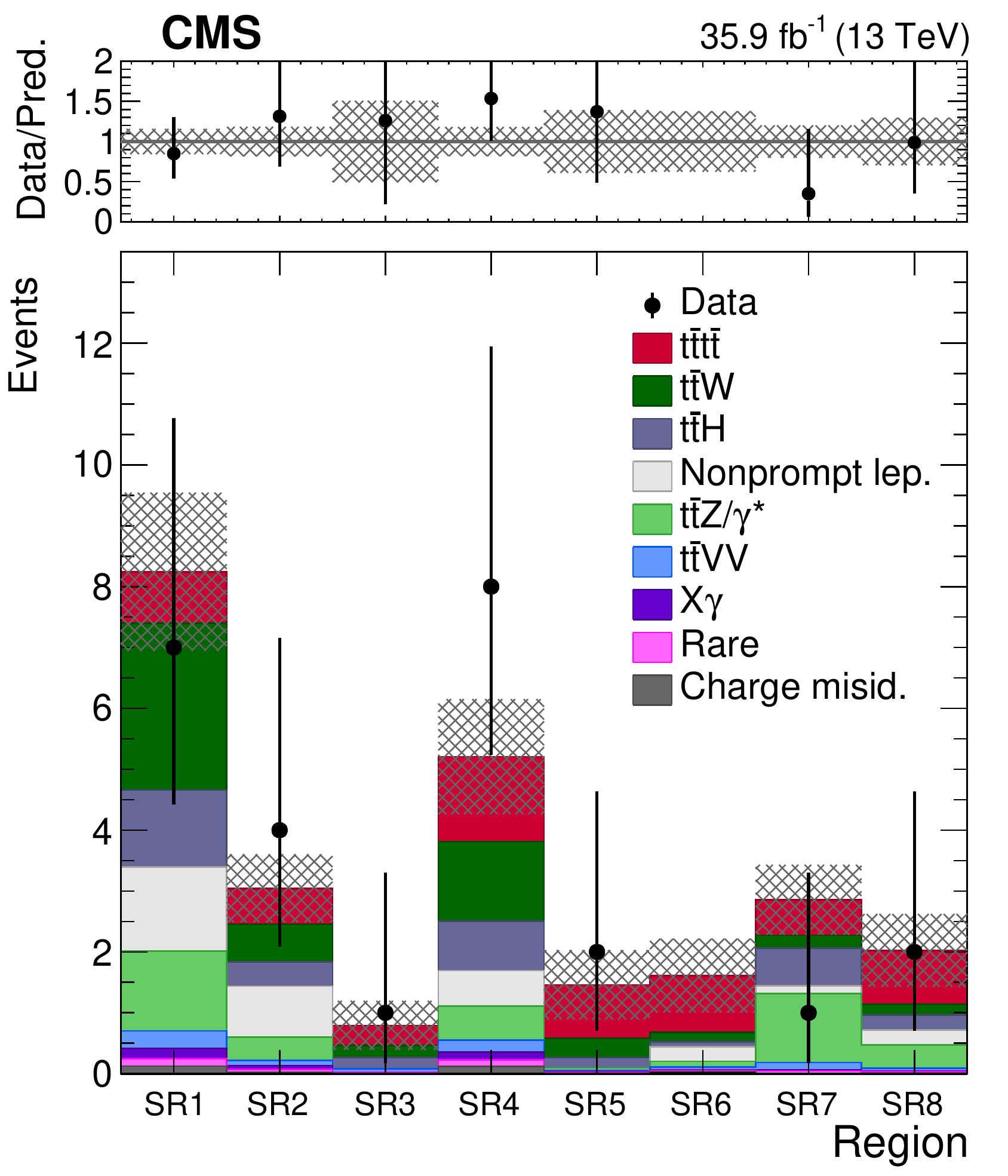}
\caption{Observed yields in the control and signal regions (\cmsLeft, in log scale), and signal regions only (\cmsRight, in linear scale),
compared to the post-fit predictions for signal and background processes.
The hatched areas represent the total uncertainties in the signal and background predictions.
The upper panels show the ratios of the observed event yield and the total prediction of signal and background.
Bins without a data point have no observed events.
    }
\label{fig:sr}
\end{figure}

The yields from SR~1--8, CRW, and CRZ are combined in a maximum-likelihood fit, following the procedures
described in Ref.~\cite{ATL-PHYS-PUB-2011-011},
to estimate a best-fit cross section for \tttt, the significance of the observation relative to the background-only hypothesis,
and the upper limit on \xsectttt.
The experimental and theoretical uncertainties described in Section~\ref{sec:systematics} are incorporated in the likelihood
as ``nuisance'' parameters and are profiled in the fit.
Nuisance parameters corresponding to systematic uncertainties are parameterized as log-normal distributions.
The fitted values of the nuisance parameters are found to be consistent with their initial values within uncertainties.
The nuisance parameters corresponding to the \ttW and \ttZgamma normalizations are scaled by $1.2\pm0.3$ and $1.3\pm0.3$, respectively, while other background
contributions including \ttH are scaled up by $1.1$ or less.
The signal and control region results after the maximum-likelihood fit (post-fit) are shown in Fig.~\ref{fig:sr}, with the fitted \tttt signal contribution
added to the background predictions, which are given in Table~\ref{tab:yields}.
The \tttt cross section is measured to be $16.9^{+13.8}_{-11.4}$\unit{fb},
where the best-fit value of the parameter and an approximate 68\% CL confidence interval are extracted following the procedure described in Sec. 3.2 of Ref.~\cite{HiggsStat}.
The observed and expected significances relative to the
background-only hypothesis are found to be 1.6 and 1.0 standard
deviations, respectively, where the expectation is based on the central value of the NLO
SM cross section of $9.2^{+2.9}_{-2.4}$\unit{fb}~\cite{MADGRAPH5}.
The observed 95\% CL upper limit on the cross section,
based on an asymptotic formulation~\cite{Cowan:2010js} of the modified frequentist CL$_\mathrm{s}$ criterion~\cite{Junk:1999kv,Read:2002hq},
is found to be 41.7\unit{fb}.  The corresponding expected upper limit,
assuming no SM \tttt contribution to the data, is $20.8^{+11.2}_{-6.9}$\unit{fb},
showing a significant improvement relative to the value of 27\unit{fb} of Ref.~\cite{SS2016}.

\begin{figure}[!hbtp]
\centering
\includegraphics[width=\cmsFigWidth]{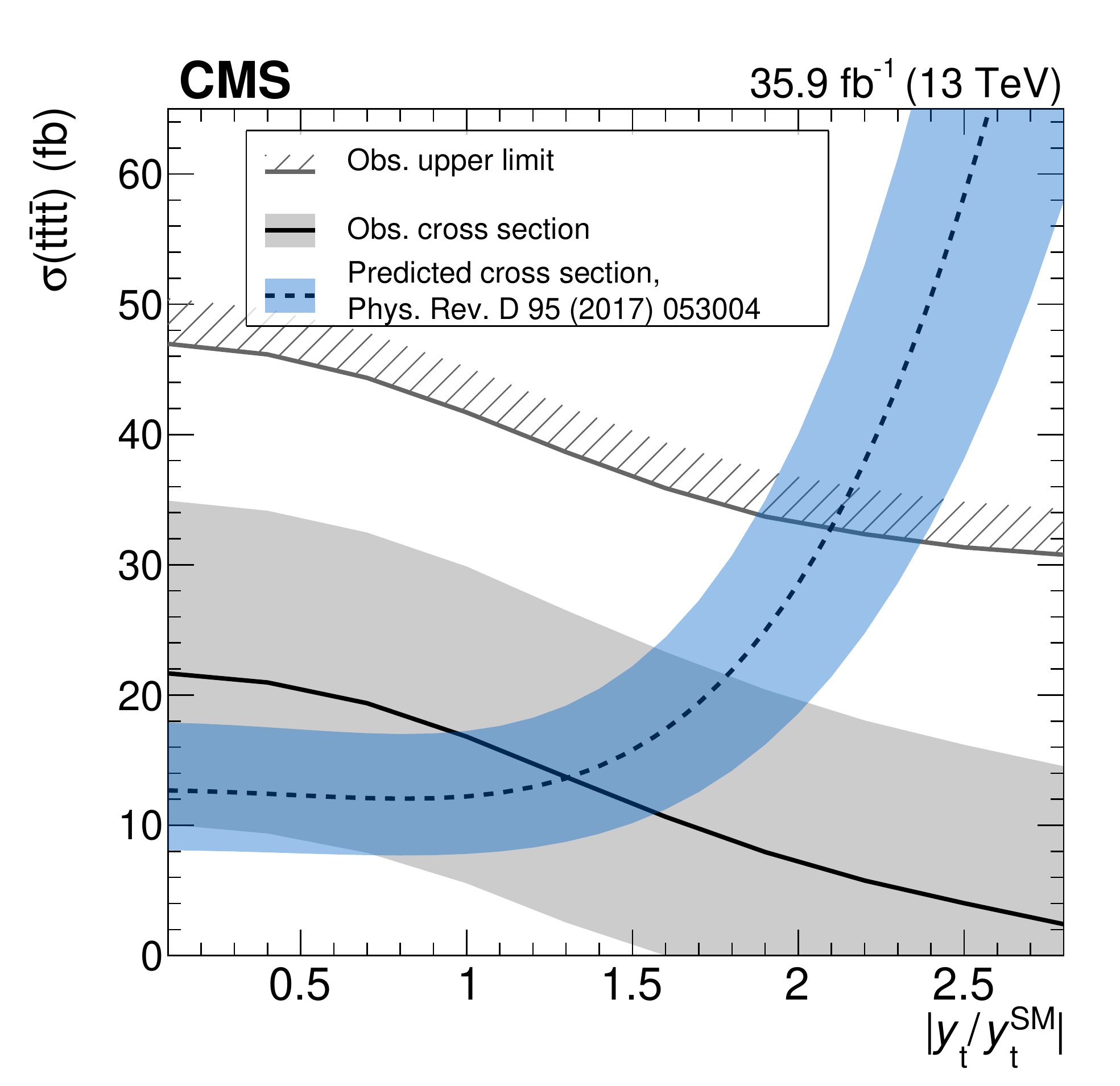}
\caption{The predicted SM value of \xsectttt~\cite{TopYukawaTTTT}, calculated at LO with an NLO/LO $K$-factor of 1.27, as a function of $\abs{y_{\PQt}/y_{\PQt}^{\mathrm{SM}}}$ (dashed line),
compared with the observed value of \xsectttt (solid line), and with the observed 95\% CL upper limit (hatched line).
}
\label{fig:yukawa}
\end{figure}

The $\Pp \Pp \to \tttt$ process has contributions from diagrams with virtual
Higgs bosons, as shown in Fig.~\ref{fig:feyndiagrams}. Experimental information on \xsectttt
can therefore be used to constrain the Yukawa coupling, $y_{\PQt}$, between the top quark and the Higgs boson.
We constrain $y_{\PQt}$ assuming that the signal acceptance is not affected by the relative contribution of the virtual Higgs boson diagrams.
As the cross section for the \ttH background also depends on the
top quark Yukawa coupling, for the purpose of constraining  $y_{\PQt}$
the fit described above is repeated with the \ttH contribution
scaled by
the square of the absolute
value of the ratio of the top quark Yukawa coupling to its SM value ($\abs{y_{\PQt}/y_{\PQt}^{\mathrm{SM}}}^2$),
where $y_{\PQt}^{\mathrm{SM}}= m_{\PQt}( \sqrt{2} G_\mathrm{F})^{1/2} \approx 1$.
This results in a
dependence of the measured \xsectttt
on $\abs{y_{\PQt}/y_{\PQt}^{\mathrm{SM}}}$ which is shown in
Fig.~\ref{fig:yukawa} and is compared to
its theoretical prediction.  The prediction is
obtained from the LO calculation of Ref.~\cite{TopYukawaTTTT}, with
an NLO/LO $K$-factor of 1.27~\cite{Bevilacqua2012}.
The LO calculation is used instead of the NLO one, as Ref.~\cite{TopYukawaTTTT} provides
a breakdown of the contributions to the cross section according to powers of $y_{\PQt}$.
The prediction also includes the uncertainty associated with varying the renormalization and factorization scales in the LO calculation by a factor of 2.
The central, upper and lower values
of the theoretical cross section provide respective 95\% CL limits  for
$\abs{y_{\PQt}/y_{\PQt}^{\mathrm{SM}}} < 2.1$, $<$1.9 and $<$2.4.

\begin{table}[ht]
\centering
\topcaption{
The post-fit background, signal, and total yields with
          their total uncertainties and the observed number of events
          in the control and signal regions in data.
}
\label{tab:yields}
    \begin{tabular}{c|cc|c|c}
        & SM background   & $\tttt$        & Total & Observed       \\
            \hline
CRZ &  31.7 $\pm$ 4.6 &  0.4 $\pm$ 0.3 & 32.1 $\pm$ 4.6 & 35 \\
CRW &  83.7 $\pm$ 8.8 &  1.9 $\pm$ 1.2 & 85.6 $\pm$ 8.6 & 86 \\
SR1 &   7.7 $\pm$ 1.2 &  0.9 $\pm$ 0.6 &  8.6 $\pm$ 1.2 & 7  \\
SR2 &   2.6 $\pm$ 0.5 &  0.6 $\pm$ 0.4 &  3.2 $\pm$ 0.6 & 4  \\
SR3 &   0.5 $\pm$ 0.3 &  0.4 $\pm$ 0.2 &  0.8 $\pm$ 0.4 & 1  \\
SR4 &   4.0 $\pm$ 0.7 &  1.4 $\pm$ 0.9 &  5.4 $\pm$ 0.9 & 8  \\
SR5 &   0.7 $\pm$ 0.2 &  0.9 $\pm$ 0.6 &  1.6 $\pm$ 0.6 & 2  \\
SR6 &   0.7 $\pm$ 0.2 &  1.0 $\pm$ 0.6 &  1.7 $\pm$ 0.6 & 0  \\
SR7 &   2.3 $\pm$ 0.5 &  0.6 $\pm$ 0.4 &  2.9 $\pm$ 0.6 & 1  \\
SR8 &   1.2 $\pm$ 0.3 &  0.9 $\pm$ 0.6 &  2.1 $\pm$ 0.6 & 2  \\
\end{tabular}
\end{table}

\section{Summary}
\label{sec:summary}

The results of a search for standard model (SM) production of \tttt
at the LHC have been presented, using
data from  $\sqrt{s} = 13\TeV$ proton-proton collisions
corresponding to an integrated luminosity of \sslumi, collected with the CMS detector in 2016.
The analysis strategy uses same-sign dilepton as well as three- (or more) lepton events, relying on jet multiplicity and jet flavor
to  define search regions that are used  to probe the \tttt process.
Combining these  regions yields a significance of 1.6 standard deviations relative to the background-only hypothesis,
 and a measured value for the \tttt cross section of $16.9^{+13.8}_{-11.4}$\unit{fb},
in agreement with the standard model predictions.
The results are also re-interpreted to constrain the
 ratio of the top quark Yukawa coupling to its
SM value, yielding $\abs{y_{\PQt}/y_{\PQt}^{\mathrm{SM}}} < 2.1$ at 95\% confidence level.

\begin{acknowledgments}
We thank Qing-Hong Cao, Shao-Long Chen, and Yandong Liu for providing calculations used in determining a constraint on the top quark Yukawa coupling.

We congratulate our colleagues in the CERN accelerator departments for the excellent performance of the LHC and thank the technical and administrative staffs at CERN and at other CMS institutes for their contributions to the success of the CMS effort. In addition, we gratefully acknowledge the computing centers and personnel of the Worldwide LHC Computing Grid for delivering so effectively the computing infrastructure essential to our analyses. Finally, we acknowledge the enduring support for the construction and operation of the LHC and the CMS detector provided by the following funding agencies: BMWFW and FWF (Austria); FNRS and FWO (Belgium); CNPq, CAPES, FAPERJ, and FAPESP (Brazil); MES (Bulgaria); CERN; CAS, MoST, and NSFC (China); COLCIENCIAS (Colombia); MSES and CSF (Croatia); RPF (Cyprus); SENESCYT (Ecuador); MoER, ERC IUT, and ERDF (Estonia); Academy of Finland, MEC, and HIP (Finland); CEA and CNRS/IN2P3 (France); BMBF, DFG, and HGF (Germany); GSRT (Greece); OTKA and NIH (Hungary); DAE and DST (India); IPM (Iran); SFI (Ireland); INFN (Italy); MSIP and NRF (Republic of Korea); LAS (Lithuania); MOE and UM (Malaysia); BUAP, CINVESTAV, CONACYT, LNS, SEP, and UASLP-FAI (Mexico); MBIE (New Zealand); PAEC (Pakistan); MSHE and NSC (Poland); FCT (Portugal); JINR (Dubna); MON, RosAtom, RAS, RFBR and RAEP (Russia); MESTD (Serbia); SEIDI, CPAN, PCTI and FEDER (Spain); Swiss Funding Agencies (Switzerland); MST (Taipei); ThEPCenter, IPST, STAR, and NSTDA (Thailand); TUBITAK and TAEK (Turkey); NASU and SFFR (Ukraine); STFC (United Kingdom); DOE and NSF (USA).

\hyphenation{Rachada-pisek} Individuals have received support from the Marie-Curie program and the European Research Council and Horizon 2020 Grant, contract No. 675440 (European Union); the Leventis Foundation; the A. P. Sloan Foundation; the Alexander von Humboldt Foundation; the Belgian Federal Science Policy Office; the Fonds pour la Formation \`a la Recherche dans l'Industrie et dans l'Agriculture (FRIA-Belgium); the Agentschap voor Innovatie door Wetenschap en Technologie (IWT-Belgium); the Ministry of Education, Youth and Sports (MEYS) of the Czech Republic; the Council of Science and Industrial Research, India; the HOMING PLUS program of the Foundation for Polish Science, cofinanced from European Union, Regional Development Fund, the Mobility Plus program of the Ministry of Science and Higher Education, the National Science Center (Poland), contracts Harmonia 2014/14/M/ST2/00428, Opus 2014/13/B/ST2/02543, 2014/15/B/ST2/03998, and 2015/19/B/ST2/02861, Sonata-bis 2012/07/E/ST2/01406; the National Priorities Research Program by Qatar National Research Fund; the Programa Severo Ochoa del Principado de Asturias; the Thalis and Aristeia programs cofinanced by EU-ESF and the Greek NSRF; the Rachadapisek Sompot Fund for Postdoctoral Fellowship, Chulalongkorn University and the Chulalongkorn Academic into Its 2nd Century Project Advancement Project (Thailand); the Welch Foundation, contract C-1845; and the Weston Havens Foundation (USA).
\end{acknowledgments}

\bibliography{auto_generated}

\cleardoublepage \appendix\section{The CMS Collaboration \label{app:collab}}\begin{sloppypar}\hyphenpenalty=5000\widowpenalty=500\clubpenalty=5000\textbf{Yerevan Physics Institute,  Yerevan,  Armenia}\\*[0pt]
A.M.~Sirunyan, A.~Tumasyan
\vskip\cmsinstskip
\textbf{Institut f\"{u}r Hochenergiephysik,  Wien,  Austria}\\*[0pt]
W.~Adam, F.~Ambrogi, E.~Asilar, T.~Bergauer, J.~Brandstetter, E.~Brondolin, M.~Dragicevic, J.~Er\"{o}, A.~Escalante Del Valle, M.~Flechl, M.~Friedl, R.~Fr\"{u}hwirth\cmsAuthorMark{1}, V.M.~Ghete, J.~Grossmann, J.~Hrubec, M.~Jeitler\cmsAuthorMark{1}, A.~K\"{o}nig, N.~Krammer, I.~Kr\"{a}tschmer, D.~Liko, T.~Madlener, I.~Mikulec, E.~Pree, N.~Rad, H.~Rohringer, J.~Schieck\cmsAuthorMark{1}, R.~Sch\"{o}fbeck, M.~Spanring, D.~Spitzbart, A.~Taurok, W.~Waltenberger, J.~Wittmann, C.-E.~Wulz\cmsAuthorMark{1}, M.~Zarucki
% ...
\vskip\cmsinstskip
% ...
1:~~Also at Vienna University of Technology, Vienna, Austria\\
% ...

\end{sloppypar}
\end{document}